# Visible absorbing TiO2 thin films by physical deposition methods

Litty Varghese, AnuradhaPatra, Biswajit Mishra, DeepaKhushalani, Achanta Venu Gopal

**Abstract**: Titanium dioxide is one of the most widely used wide bandgap materials. However, the $TiO_2$ deposited on a substrate is not always transparent leading to a loss in efficiency of the device, especially, the photo response. Herein, we show that atomic layer deposition (ALD) and sputtered $TiO_2$ thin films can be highly absorbing in the visible region. While in ALD, the mechanism is purported to be due to oxygen deficiency, intriguingly, in sputtered films it has been observed that in fact oxygen rich atmosphere leads to visible absorption. We show that the oxygen content during deposition, the resistivity of the film could be controlled and also the photocatalysis response has been evaluated for both the ALD and sputtered films. High resolution TEM and STEM studies show that the origin of visible absorption could be due to the presence of nanoparticles with surface defects inside the amorphous film.

## Introduction

Titanium dioxide is one of the most commonly used and studied semiconductors in the field of optics, solar cells, integrated optoelectronics and environmental issues due to its high refractive index, excellent optical transmission,large energy band gap, high dielectric constant, very good wear resistance, along with a high chemical resistance against solvents [1-10]. Photocatalytic behavior of $TiO_2$ in UV region has been extensively studied [11-17]. Under UV irradiation, the decomposition of organic compounds has been easily facilitated by $TiO_2$ and it is increasingly being established that the surface of $TiO_2$can be used for self-cleaning and anti-fogging effects [18]. The principle behind photocatalytic reactions involving $TiO_2$ is that though the photoexcited carriers predominantly recombine in a non-radiative manner, few of them diffuse to the surface and take part in a variety of redox reactions such as generation of hydrogen and oxygen (Eq 1), or form different reactive oxygen species (*e.g.* $OH^{\cdot}$) through simultaneous reduction and oxidation of water or molecular oxygen (Fig. 1).

$$2H_2O\,(l) + 4h^+ = O_2\,(g) + 4H^+\ [E^0 = 1.23\ eV] \qquad \text{Eq 1a}$$

$$4H^+ + 4e^- = 2H_2\,(g)\ [E^0 = 0\ eV] \qquad \text{Eq 1b}$$

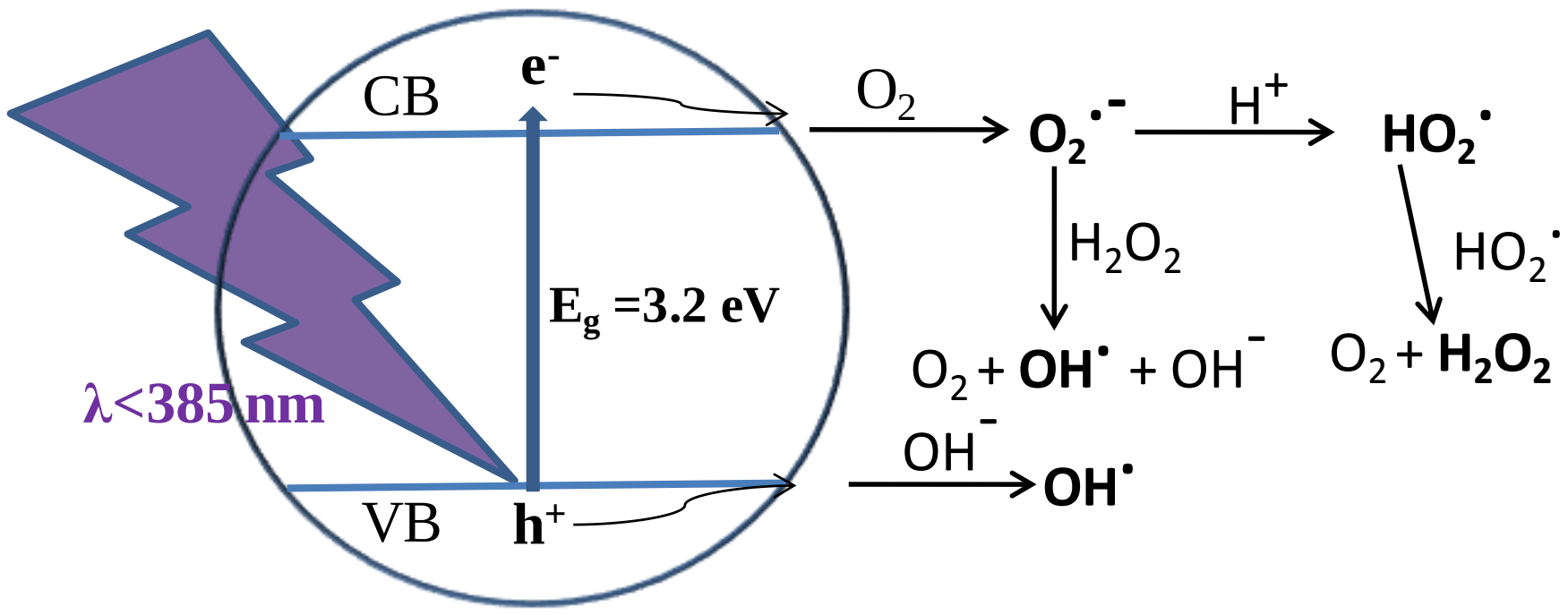

Fig. 1 Schematic representation of the main processes involved in photocatalysis

If any organic pollutant is present, the formed reactive oxygen species react with the pollutants adsorbed on the surface of $TiO_2$ and ideally mineralize them to $H_2O$ and $CO_2$. Though this criterion could also be satisfied with a number of other semiconductors [ref] (Reference: A. Ghicov, P. Schmuki, Chem. Commun., 2009, 2791.), the matching of the band edges with the important half reactions to generate reactive oxygen species make $TiO_2$ a special material as a photocatalyst.

However, due to energy gap of ~3eV, its efficiencyin absorbing sunlight is minimal as radiation from Sun consists only 5% UV while, predominant, 43% is in the visible and 52% is in infraredregion [19]. To make use of solar energy, viable efforts for reducing the bandgap of $TiO_2$ were started in 1990s. These methods have been mainly based on variation of $TiO_2$ stoichiometry by doping usinga variety of elements [20-21]. As a result of low lying dopant states, improved photo-catalytic activity has been reported [22-23]. Recently, defect engineered black $TiO_2$ nanoparticles [24] and nanowires [25] by hydrogenation have been reported and this has led invariably to enhanced visible absorption and photocatalytic activity. Subsequently several other reports have contributed to this area with detailed insight being gained into the manner in which hydrogen incorporation leads to a decrease in the band gap [26]. The development of reduced form of $TiO_2$ by oxygen vacancies as a defect also resulted in visible absorption [27-29] which has improved photo catalytic activity and in some cases splitting of water to form hydrogen has also been observed.

It is generally accepted that $TiO_2$ exhibits three crystalline phases. Two are tetragonal phases: rutile and anatase and the third one, brookite, is orthorhombic. Depending on the deposition conditions any of these phases can be realized. Among them rutile is the most thermodynamically stable phase. However anatase phase is known to have better electron mobility. As a result, predominantly this phase has been preferred for photocatalysis. Band structure calculations revealed that rutile and anatase have a direct and indirect band gap, respectively. Although wide range of band-gap energies have been reported for both rutile and anatasephase $TiO_2$ by optical measurements, anatase is known to have higher band-gap energy [31-37].

Techniques used for growing $TiO_2$ films include magnetron sputtering [38, 39], sol-gel method [40], ion-assisted deposition [41,42], pulsed laser deposition [43], molecular-beam epitaxy [44] and atomic layer deposition (ALD) [45-46]. In this paper we present preparation of visible absorbing $TiO_2$ thin films having exclusively rutile phase using two physical vapour deposition techniques:(1) radio frequency (RF)sputtering and (2) atomic layer deposition (ALD). These techniques have been known to be suitable for realizing thin films with controlled structure, composition, desired phase and uniform coating as there is control over the deposition parameters such as temperature, rate and the ambient atmosphere. The thin films have been characterized by x-ray diffraction, white light transmission, resistivity, scanning electron microscope (SEM), Transmission electron microscopy (TEM) and studied their photo-catalytic properties.

For ALD, Cambridge Savannah 100 Atomic Layer Deposition System was used for the preparation of uniformly coated thin films. Thin films were deposited on quartz and silicon at the growth temperature of $200^oC$ (Fig.2). Titanium tetrachloride ($TiCl_4$) from Sigma-Aldrich

Chemicals Ltd was used as the source for the growth of $TiO_2$ with $N_2$as the carrier gas. The base pressure was $10^{-5}$ Torr and operating pressure was 50 mTorr. Pure $H_2O$ was used as oxygen precursor. Prevention of gas-phase reactions of the vapors was achieved by evacuation of chamber to a pressure below $10^{-4}$Torr between successive pulses. Repetition of requisite number of cycles consisting of alternating $H_2O$ and Ti pulses resulted in the formation of $TiO_2$ thin film growth layer by layer with atomic precision. The rate of $TiO_2$ deposition was 0.4Å/cycle. In the second method, sputtered thin films were realized using a confocal sputter system. The base pressure was $4x10^{-6}$ mbar and operating pressure was $2x10^{-2}$ mbar in Ar plasma. The RF power was 100 W at room temperature with a rotating stage. Ar and $O_2$ gas combination was used. For Ar:$O_2$ ratio of 70:30 sccm, the resulting film was dark. The rate of deposition was 0.67 nm/min.

The thickness of thin films was measured using Ambios profilometer and Sentech's Film Thickness probe (FTP) which works in the reflection mode. The SEM studies were done using a ZEISS Ultra 55 plus FESEM. For AFM studies, Park XE70 with Silicon cantilever in non-contact mode was employed and root mean square value (RMS) for roughness was calculated using commercial XEI software. X-ray Diffraction (XRD) measurements were done using X'pert pro XRD with Cu Kα ($\lambda$ = 1.54056 Å) source. Data was collected in glancing angle mode with source angle at 1° and analyzer angle varied from 20° to 60° with a step size of 0.025(2Θ). X-ray Photoelectron Spectroscopy (XPS) was carried out using ESCA from Omicron Nanotechnology. Al Kα monochromatic source was used with a hemispherical analyzer. FEI TITAN transmission electron microscope (TEM) was used for transmission electron microscopy studies. The four probe resistivity and photovoltaic studies were done using Signatone 4-probe station, Keithley source-meter and Lock-in amplifier at room temperature. Photoconductive studies were done with the probe station and Lock-in amplifier combination in the presence of white light source (100 W Halogen lamp). Transmission studies were carried out using a Cary 5000 spectrophotometer in the 200 nm to 2500 nm range. Reflection studies were done with Jasco spectrophotometer covering the 200 to 2500 nm wavelength range. Refractive index studies were done using Sentech 800 ellipsometer.

The visible image showed that film prepared by ALD was black in color and sputtered ones were dark blue in color. From AFM data, the surface roughness could be obtained to be 1.3 nm for ALD grown film and 0.8 nm for sputtered films (Fig.3a and b). Optical transmission and reflection studies revealed 30% absorption in the visible region for the samples prepared by sputtering in the presence of Ar/$O_2$ compared to the film prepared in the presence of solely Argon (Fig.3a and b). Transmission studies on ALD thin film showed a 50% reduction in transmitted light in the visible region compared to $TiO_2$ deposited in Ar atmosphere (Fig.4a and b). Ellipsometric measurements on thin films showed that refractive index of sputtered thin film was 2.55 and that of ALD film was 2.7 at 500 nm wavelength (Fig.5).

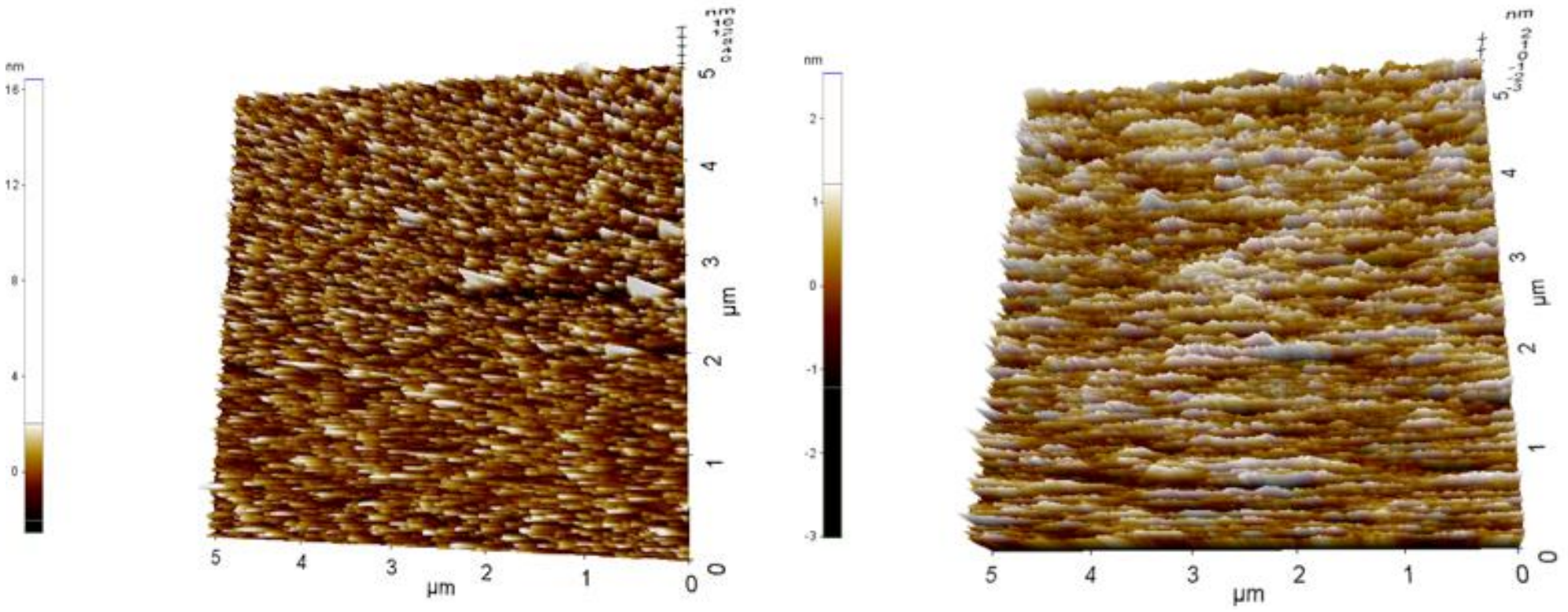

**Fig.2. AFM image of $TiO_2$ thin film prepared a) by ALD and b) by sputtering**

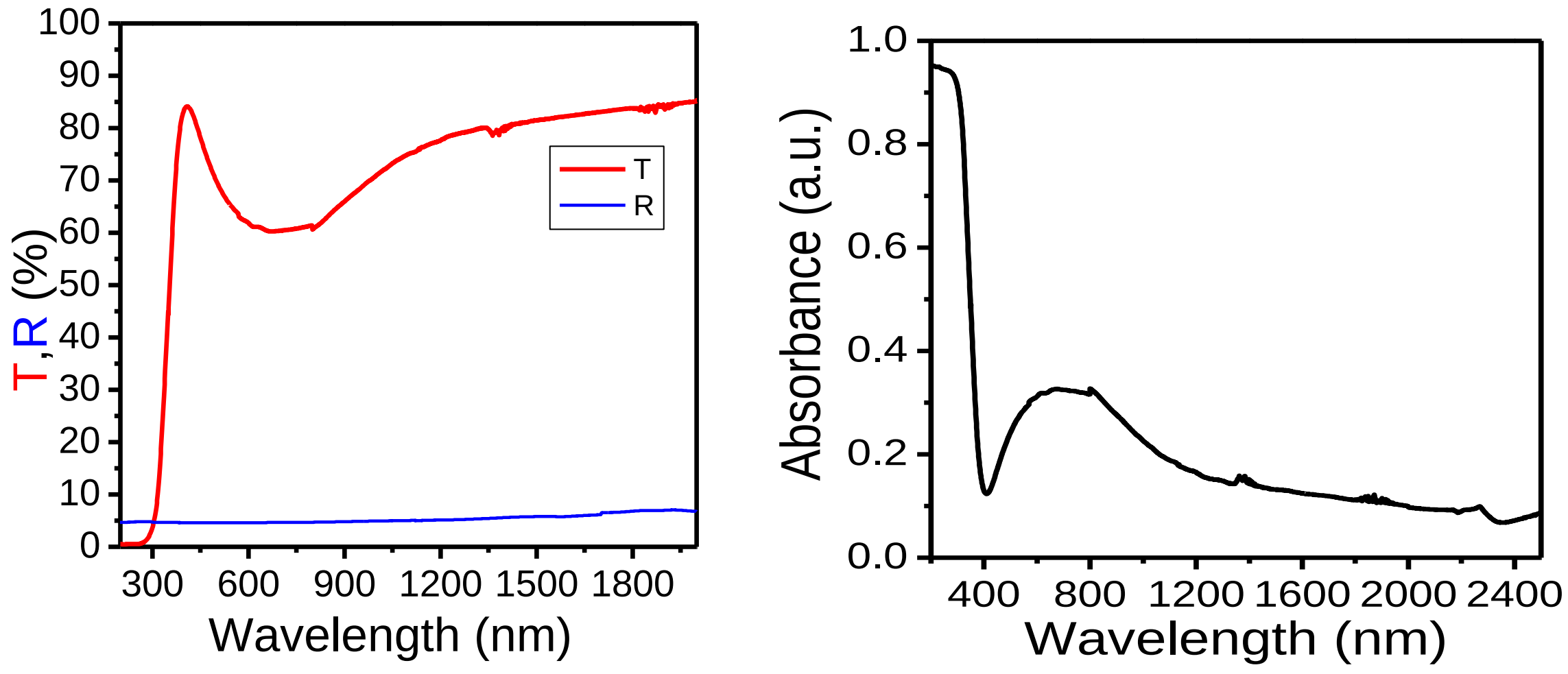

Fig.3. Transmission and reflection studies on sputtered $TiO_2$ thin film a) in presence of Argon, b) in presence of Ar/$O_2$ and c) absorption spectrum of sputtered thin film in presence of Ar/O2

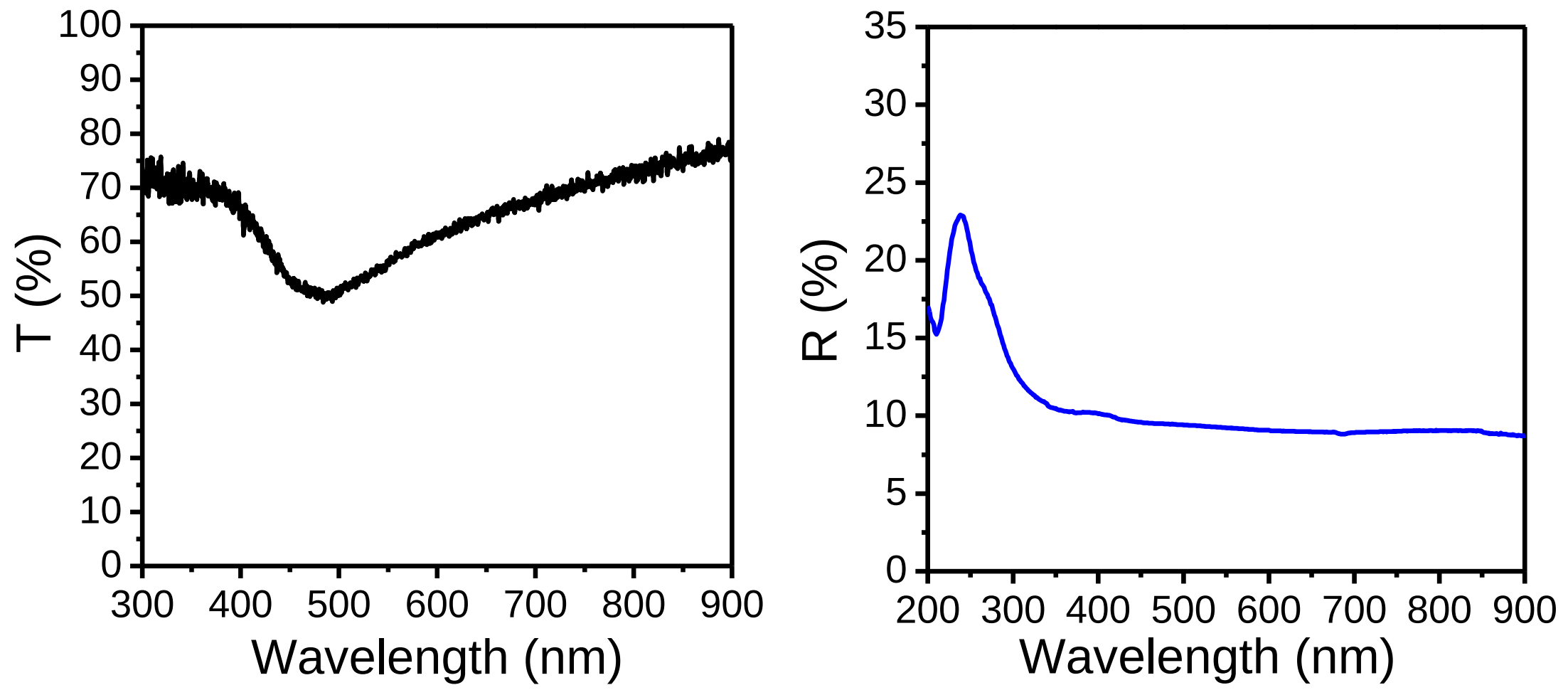

**Fig: 4. a) Transmission and b) reflection studies on ALD grown $TiO_2$ thin film**

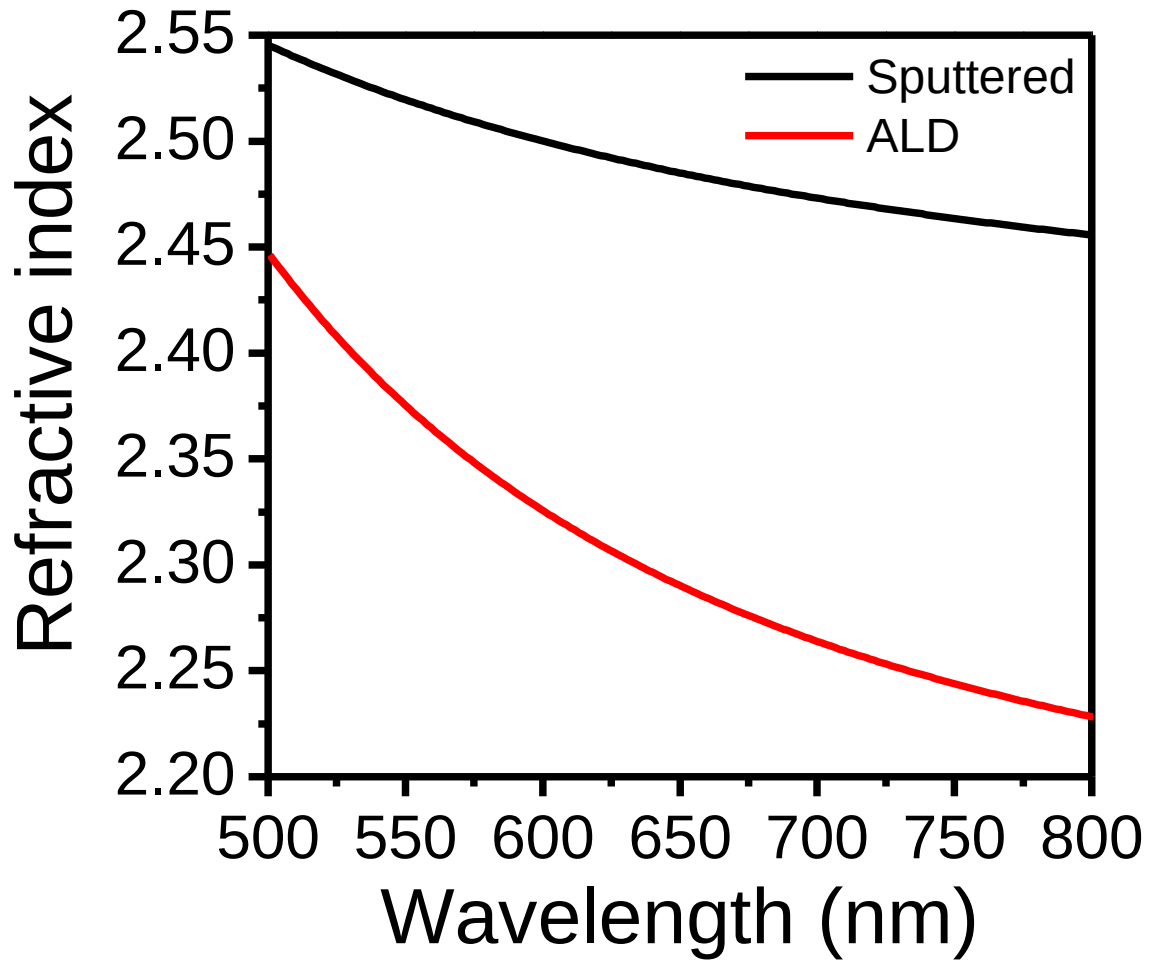

**Fig.5. Dispersion measurements on TiO2 thinfilms**

High resolution Transmission electron microscopy (HRTEM) and Scanning electron microscope (SEM) studies done on these films (Fig.6) showed the presence of highly crystalline defect induced individual $TiO_2$ nanoparticles of around 3 nm in diameter. Clusters of $TiO_2$ nanoparticles with highly resolved lattice features are found in amorphous matrix of $TiO_2$ as shown in Fig.7. Strong peaks obtained from XRD studies showed that highly crystalline $TiO_2$ of rutile phase in both the thin films (Fig.8). The crystallite size calculated from XRD peaks were 29 nm which seems to be in good agreement with the size of clusters shown in TEM.

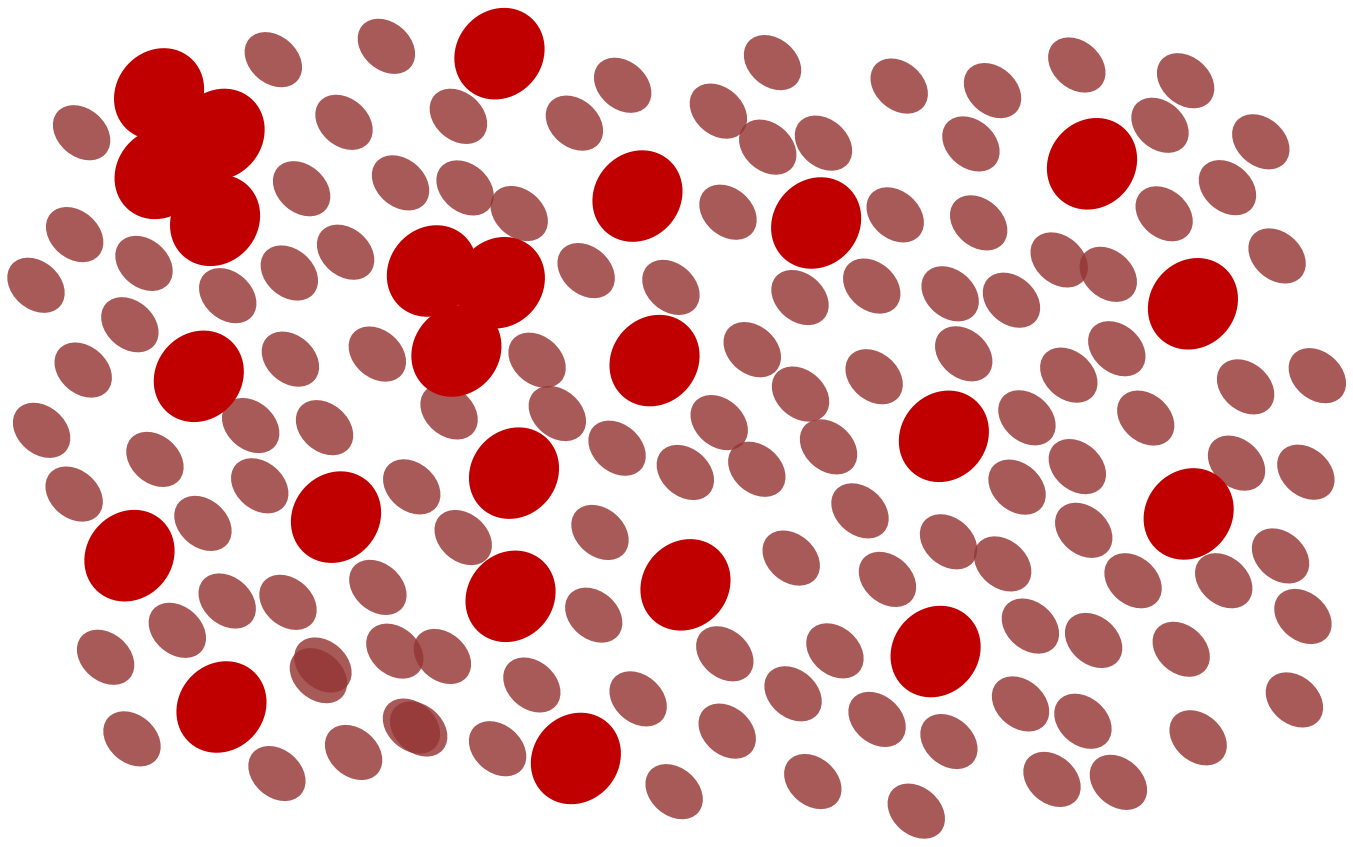

Fig.7. Schematic view of TiO2 nanoparticles embedded in amorphous matrix

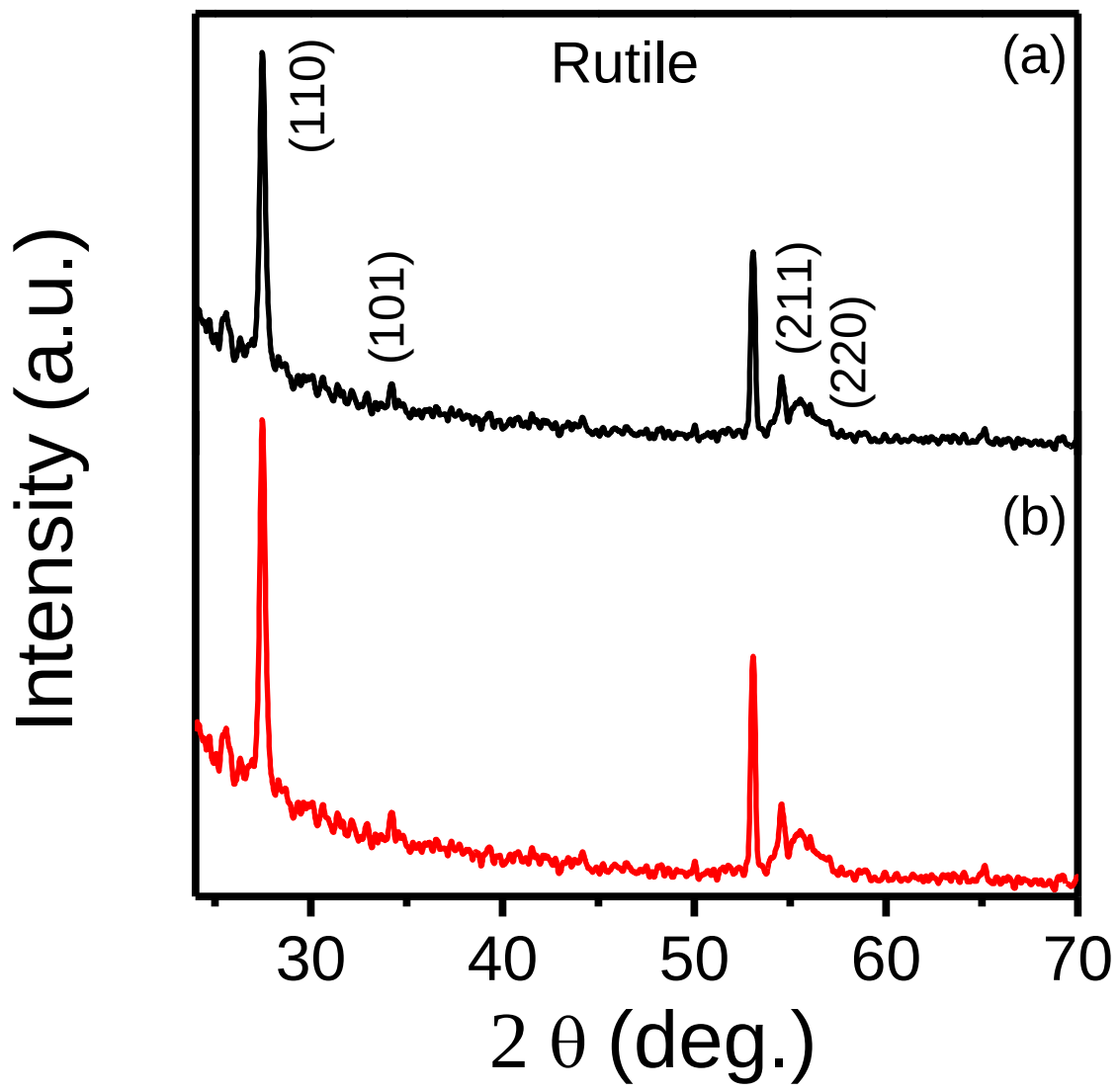

Fig.8. XRD studies on thin films prepared by a) ALD and b) sputtering

In order to gain further insight into the nature of the films and the electronic environment around Ti and O centers, XPS measurements were performed. The Ti 2p core level spectrum of $TiO_2$ thin film grown by ALD exhibits symmetric spin orbit split lines at 458.224 eV and 463.999 eV for Ti $2p_{3/2}$ and Ti $2p_{1/2}$, respectively, with a difference of 5.7 eV between the doublets (Fig. 8). The O 1s region showed a peak at 529.738 eV, characteristic of oxides of transition metals and a prominent shoulder at 531.523 eV which is a signature of the presence of disorder in the surface in the form of adsorbed oxygen [47,48]. The peak at 531.523 eV is designated as higher binding energy peak (HBEP) in Fig. 8b. In addition, a broad peak centered at ~400 eV (around N 1s region) is observed and is attributed to the molecularly chemisorbed nitrogen (γ-$N_2$) (Fig.9). However, no trace of atomic nitrogen (β-N, Ti-N bonding) is observed in the XPS spectrum. The O 1s peak in samples annealed in presence of $N_2$ showed a shift of 0.577 eV towards lower binding energy and did exhibit a minor change in intensity. The result indicates that the as-grown ALD samples are oxygen deficient.

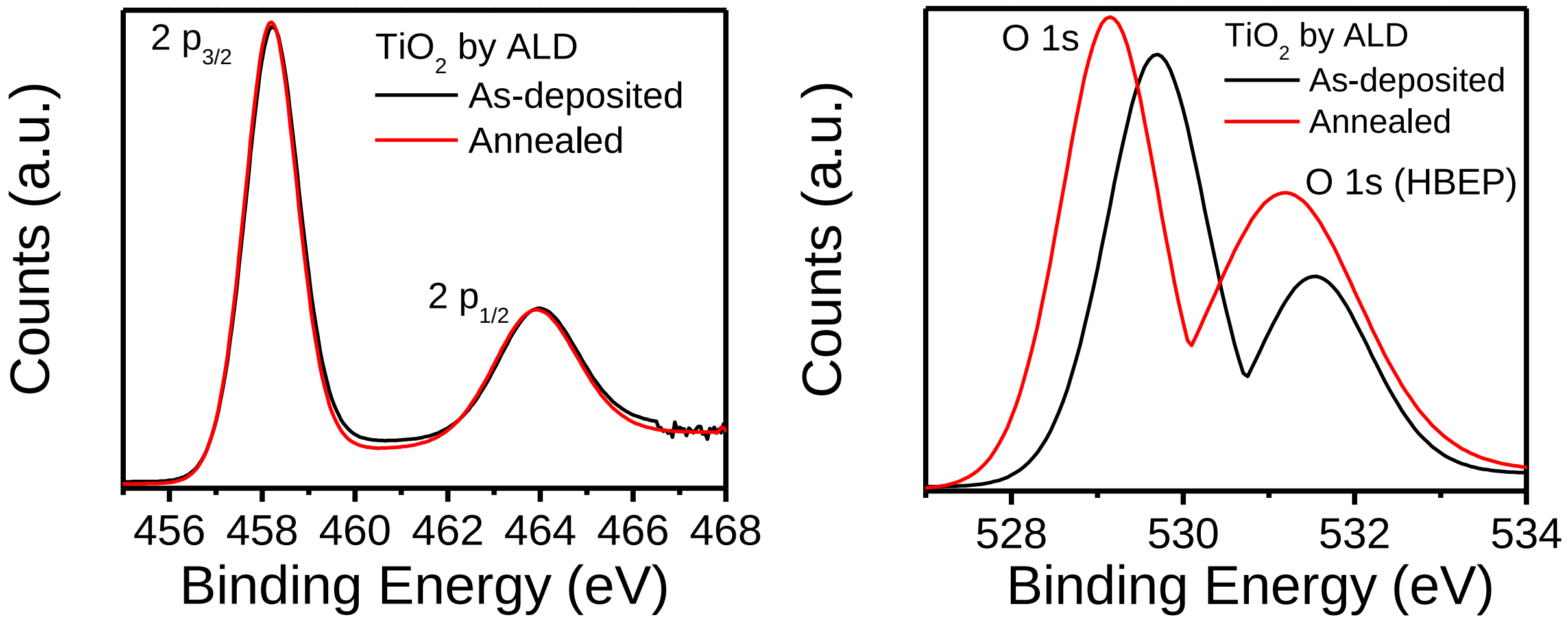

**Fig. 9 (a) XPS Ti 2p spectrum (b) O 1s spectrum of $TiO_2$ film grown by ALD**

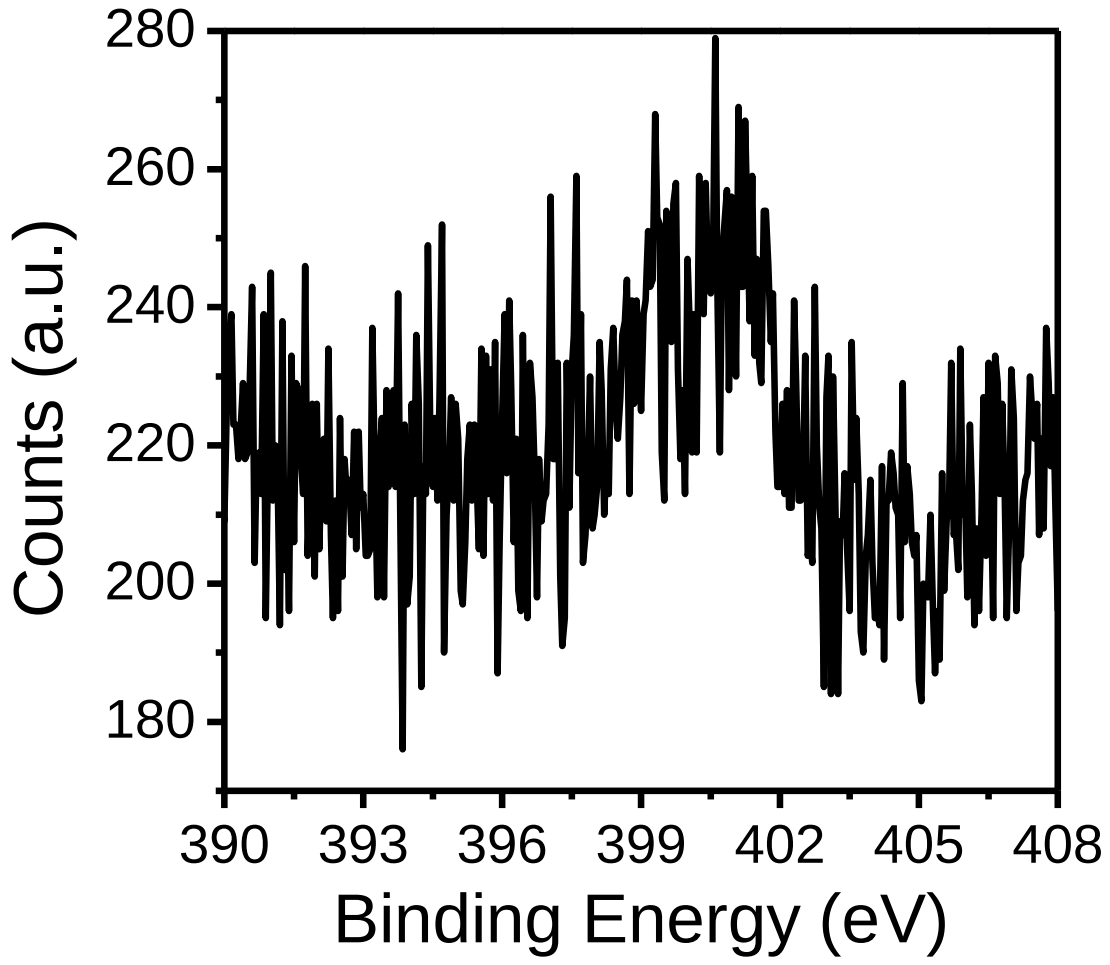

**Fig. 10 XPS spectrum of ALD grown $TiO_2$ around N 1s binding energy.**

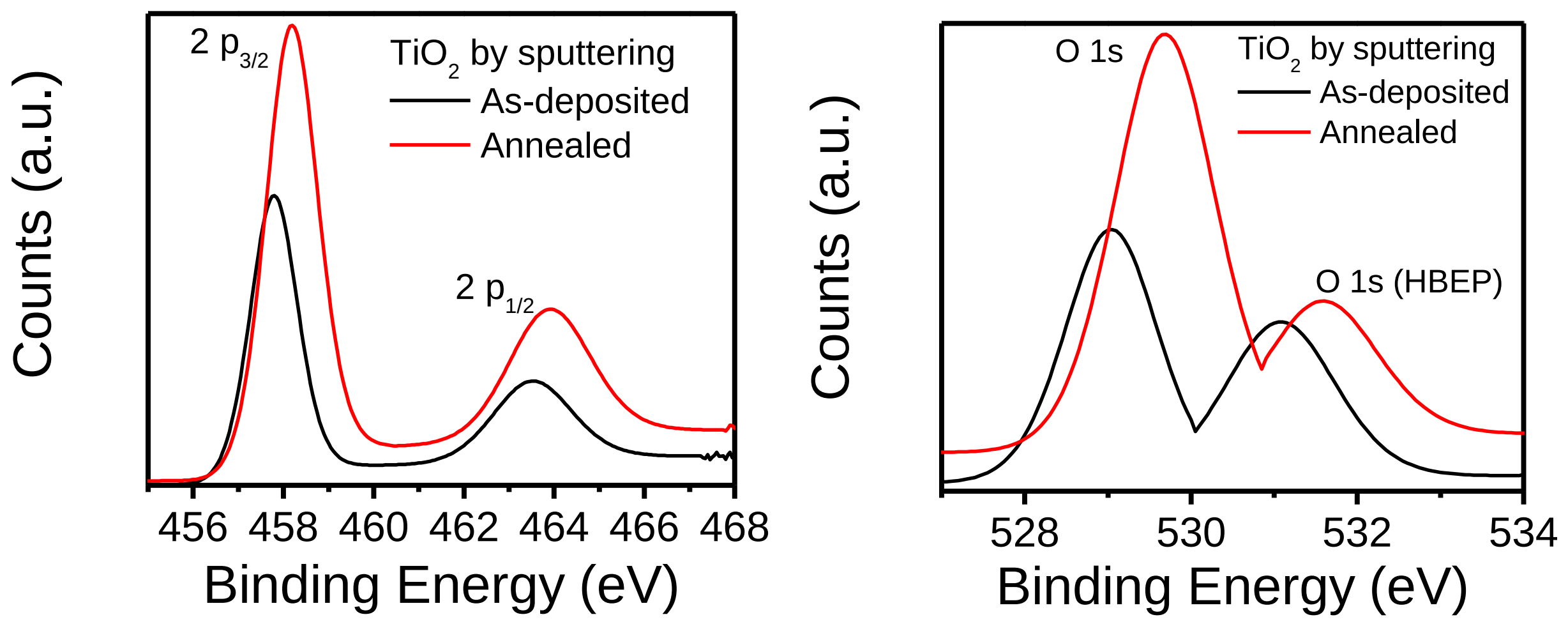

**Fig.11 XPS spectra of (a) Ti 2p and (b) O 1s of sputtered $TiO_2$ thin film**

The XPS spectrum of $TiO_2$ thin film deposited by sputtering in the presence of Argon and oxygen plasma is plotted in Fig.10. The Ti $2p_{3/2}$, Ti $2p_{1/2}$ and O 1s core levels are located at 457.825 eV, 463.544 eV and at 529.068 eV, respectively. The 2p peaks are symmetric, showed binding energy separation of 5.7 eV and could be fitted with a single Gaussian peak. In addition to the signature of lattice oxygen of $TiO_2$ at 529.068 eV, the O 1s region exhibits a pronounced shoulder at higher binding energy at 531.074 eV. The shoulder peak at higher binding energy has been observed by many research groups [47,48] and is generally attributed to defective oxides or sub-stoichiometric oxide on the surface of $TiO_2$. The sputtered $TiO_2$ films were subjected to heat treatment and the XPS spectrum of the annealed films is also shown in Fig. 10. It is observed that the Ti 2p peaks as well as O 1s peak shift to higher binding energy and there is a considerable enhancement in the recorded signal. Thus, the as-grown sputtered samples are oxygen rich. The Ti-O-Ti bonding strength, however, is stronger in the annealed samples as is evident from the increased signal strength. $TiO_2$ being a wide band gap semiconductor ($E_g$ ~ 3.2 eV) is transparent to visible light. However, surface defects may play a significant role in altering the optical response of the films and may lead to visible light absorption [49]. These surface defects, when present in good number introduce mid-gap states within the forbidden region. The formed continuum merges with conduction/valence band edge thus modulating optical behavior. Hence we can conclude from the XPS studies that the presence of excess oxygen results in the visible absorption in sputtered $TiO_2$ thin film which is also supported by a recent report [25].

In addition to structural and composition studies, we also conducted studies on electrical properties of $TiO_2$ thin films using four probe measurements. The measured resistivity was in kΩ-cm for ALD film and in GΩ-cm for sputtered films. For temperature dependent studies, annealing was done on samples at 400°C in atmospheric air for 20 min. The resistivity versus temperature graph (Fig.11a) suggests that the behavior of ALD thin film varies from metal to insulator (MOI) as temperature increases. The resistivity of the sputtered thin film reduces from GΩ-cm to kΩ-cm as the temperature is increased (Fig. 11b). As was inferred from the XPS analysis, the sputtered $TiO_2$ films have large amounts of surface defects (presence of excess oxygen). Due to which the resistivity of the film is very high (GΩ-cm). When the film is annealed some of the surface defects may get relaxed and consequently may lead to better electron transport which otherwise was hindered due to the presence of defects. As a result, the resistivity of the sample shows systematic decrease (Fig. 11b) with increase in temperature. The temperature dependent resistivity behavior of ALD grown $TiO_2$ film is different from that of sputtered one. The ALD film shows a decrease in measured resistivity up to 100°C and then a monotonous increase. Although the structure of $TiO_2$ films, irrespective of growth mode (sputtering or ALD), is rutile, the difference in resistivity behavior in the films suggest that there is a considerable difference in surface chemistry. In ALD grown films the chemisorbed nitrogen, owing to low sticking probability, might get desorbed on heat treatment and this may lead to initial decrease in resistivity. However, soon the surface sites are occupied with the chemisorbed oxygen, as it was observed in XPS spectrum of ALD grown $TiO_2$ film (Fig.10b) that the high energy O 1s peak intensity increases significantly on annealing. Thus the surface is densely filled with chemisorbed oxygen which may result in increased resistivity of the film with rise in annealing temperature.

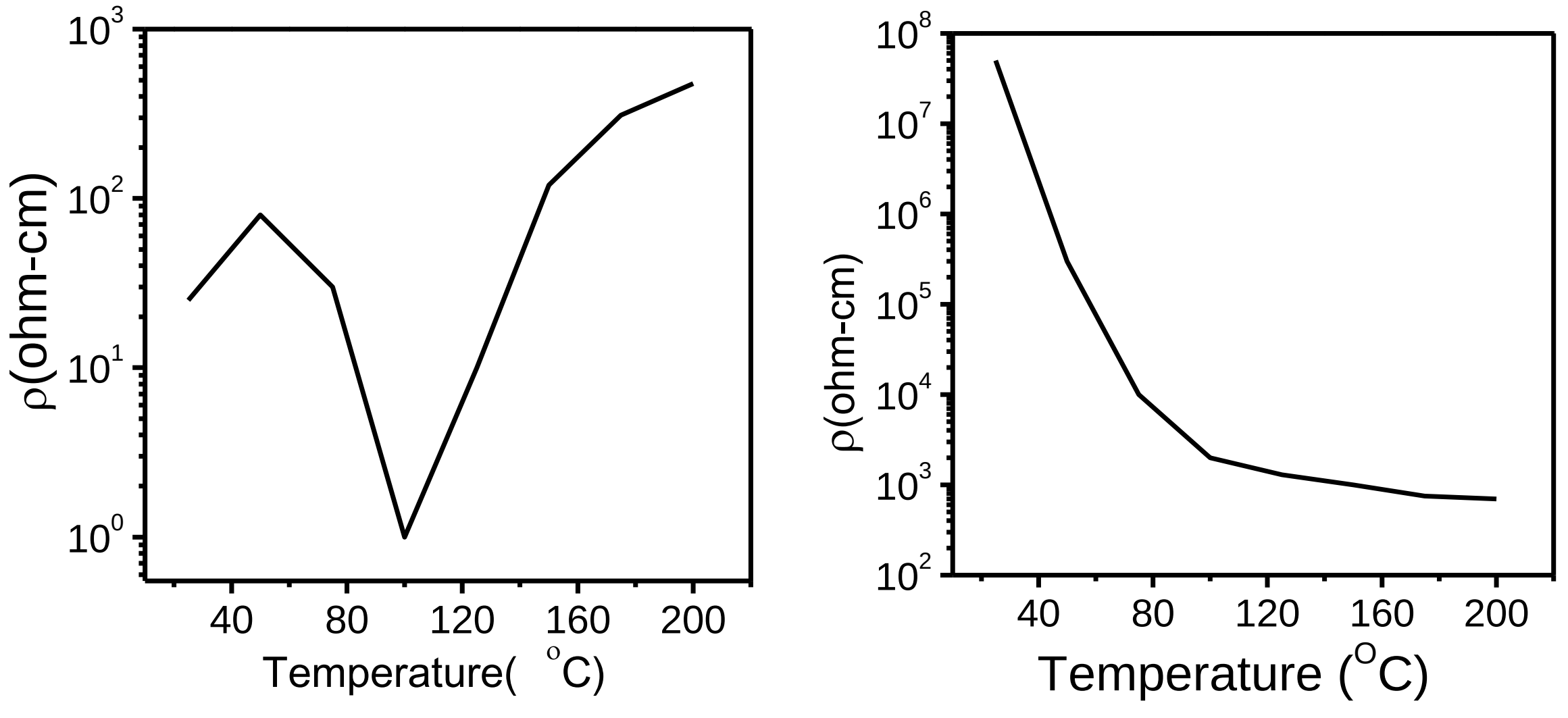

**Fig.12 Resistivity studies on a) ALD $TiO_2$ thin film and b) sputtered thin film**

As a further to evaluate the effectiveness of these films, photocatalysis experiments were conducted by using directly the films (mounted on the substrates) and suspending them in a solution of a model organic compound (tryptophan). The main impetus for photocatalysis was to evaluate the films (as they have impressive absorption in the visible range) for viability in forming long lived electrons/holes that have strong oxidative/reductive capability. As such, the films were suspended in a solution of tryptophan and the samples were carefully irradiated with exclusively visible light (long pass filter at 400 nm was used) and the degradation of the model organic compound was monitored by spectrophotometer. Note that the model compound, tryptophan has absorption only in the UV region and as such sensitization artefacts (that lead anomalously to enhanced photocatalysis rates) have been carefully avoided. Shown in Fig 12a is the spectra of tryptophan as a function of time, Fig. 12b shows a comparative plot for the degradation rate of Tryptophan in the presence of a conventional $TiO_2$ catalyst (P25, commercially available) along with the ALD prepared thin film sample. It can be readily observed that the rate of degradation of model compound is faster in the presence of the sputtered $TiO_2$ thin film sample and in fact even the ALD deposited film shows a faster rate than the conventional, oft-cited, standard catalyst P25. This result corroborates the XPS and resistivity data whereby the presence of excess oxygen species in the thin films, leads to longer lifetimes of the excited carriers. These long lived species as a result are able to diffuse readily to the surface and contribute to the formation of large ROS (reactive oxygen species) which are vital for the degradation of the model organic compound.

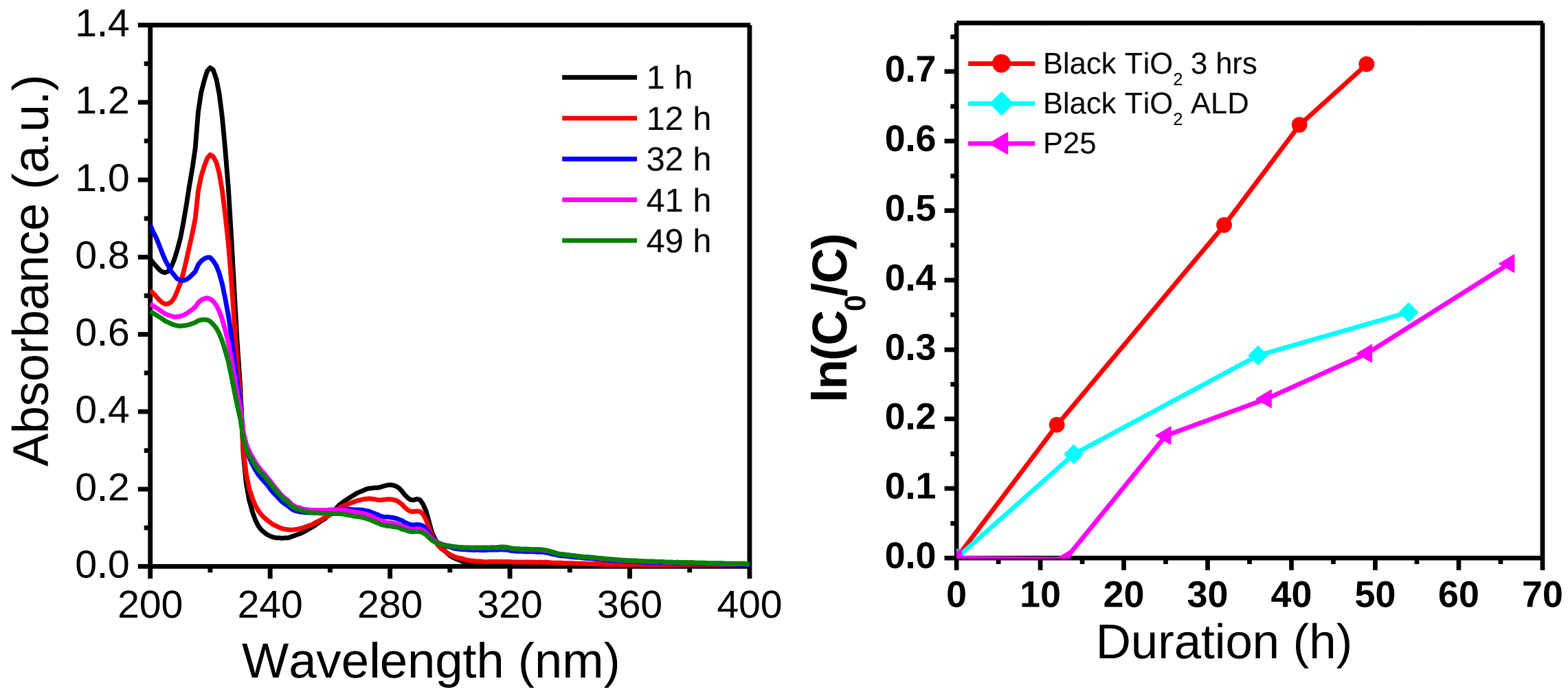

**Fig. 13 Caption**